\documentclass[
aps,
prl,
twoside,
twocolumn,
10pt,
floatfix,
showpacs,
citeautoscript,
superscriptaddress,
]{revtex4-2}

\usepackage{amsmath}
\usepackage{amssymb}
\usepackage{glossaries}
\usepackage{graphicx}
\usepackage[version=3]{mhchem}
\usepackage[per-mode = symbol, range-units=single, range-phrase=--]{siunitx}

\usepackage[colorlinks=true, urlcolor=black, linkcolor=black, citecolor=black]{hyperref}

\graphicspath{{figures/}}

\newacronym{bo}{BO}{Bayesian optimization}
\newacronym{df}{DF}{dielectric function}
\newacronym{fdtd}{FDTD}{finite-difference time-domain}
\newacronym[longplural={figures of merit}]{fom}{FoM}{figure of merit}
\newacronym{pmma}{PMMA}{poly(methyl methacrylate)}
\newacronym{slr}{SLR}{surface lattice resonance}

\DeclareSIUnit\sccm{\text{sccm}}  
\DeclareSIUnit\torr{\text{Torr}}
\DeclareSIUnit\rpm{\text{rpm}}
\DeclareSIUnit\bar{\text{bar}}

\makeatletter
\let\oldtheequation\theequation
\renewcommand\tagform@[1]{\maketag@@@{\ignorespaces#1\unskip\@@italiccorr}}
\renewcommand\theequation{(\oldtheequation)}
\makeatother

\newcommand{\phys}{%
    Department of Physics and Astronomy,
    Chalmers University of Technology,
    SE-412~96 Gothenburg, Sweden%
}
\newcommand{\vua}{%
    Department of Physics and Astronomy,
    Vrije Universiteit Amsterdam, De Boelelaan 1100, 1081 HZ Amsterdam, Netherlands%
}
\newcommand{\wise}{
    Wallenberg Initiative Materials Science for Sustainability,
    Chalmers University of Technology,
    SE-412~96 Gothenburg, Sweden%
}

\begin{document}

\title{Bayesian Optimization for Practical \texorpdfstring{\ce{H2}}{H2} Sensors: \texorpdfstring{\\}{}Inverse Design of Pd-based Plasmonic Metasurfaces}

\author{Pernilla Ekborg-Tanner}
\affiliation{\phys}
\author{Athanasios Theodoridis}
\affiliation{\phys}
\author{Joachim Fritzsche}
\affiliation{\phys}
\author{Christoph Langhammer}
\affiliation{\phys}
\author{Andrea Baldi}
\affiliation{\vua}
\author{Paul Erhart}
\affiliation{\phys}
\affiliation{\wise}
\email{erhart@chalmers.se}

\begin{abstract}
Hydrogen detection is becoming increasingly important as its use grows across energy and industrial systems.
Optical sensing platforms based on palladium (Pd) nanoparticles are attractive for this task because hydrogen uptake directly alters their plasmonic response.
Organizing such nanoparticles into periodic two-dimensional arrays, known as metasurfaces, further enhances their optical response through collective resonances.
However, the large design space presented by chemical composition, nanoparticle geometry, and array structure calls for systematic approaches for optimizing complex nanoalloy metasurface geometries.
Here, we develop an inverse-design framework based on Bayesian optimization that couples first-principles dielectric functions with electromagnetic simulations to identify high-performance PdAu nanodisk arrays for hydrogen sensing in the \qtyrange{1}{100}{\milli \bar} range where the flammability of \ce{H2} becomes a concern.
We use our approach to search a five-dimensional design space, comprising nanodisk height and radius, array pitch, polymer coating thickness, and Au fraction in order to maximize the H-induced change in extinction at a single wavelength of choice.
The results show that integrating first-principles optical models with data-efficient optimization yields experimentally feasible nanoparticle metasurfaces tailored for targeted hydrogen pressures, while providing a pathway to future multiobjective sensor design.
They also reveal remaining gaps in the modeling methodologies that still limit the quantitative reliability of the approach.
\end{abstract}

\maketitle

In the search for sustainable alternatives to fossil fuels, hydrogen gas (\ce{H2}) has emerged as a promising energy carrier.
To achieve a safe hydrogen economy, efficient and reliable \ce{H2} sensors are essential due to the high flammability of \ce{H2} under ambient conditions.
Over the past decades, nanostructured palladium (Pd)-based materials have shown great promise for this purpose, providing a spark-free, optical sensing platform \cite{LanZorKas07, LanLarKas10, LiuTanHen11, BoeBanSet17, NugDarZhd18, SheSheWan19, NugDarCus19, BenYamKur19, SteStrBot20, BanSchDam21}.
This functionality is enabled by the ability of Pd to quickly absorb and release hydrogen, which simultaneously changes its optical properties.
In Pd nanodisks, this effect manifests as a shift of the localized surface plasmon resonance upon \ce{H2} exposure, which can be utilized as a sensing mechanism \cite{LanZorKas07}.

\begin{figure*}
\centering
\includegraphics{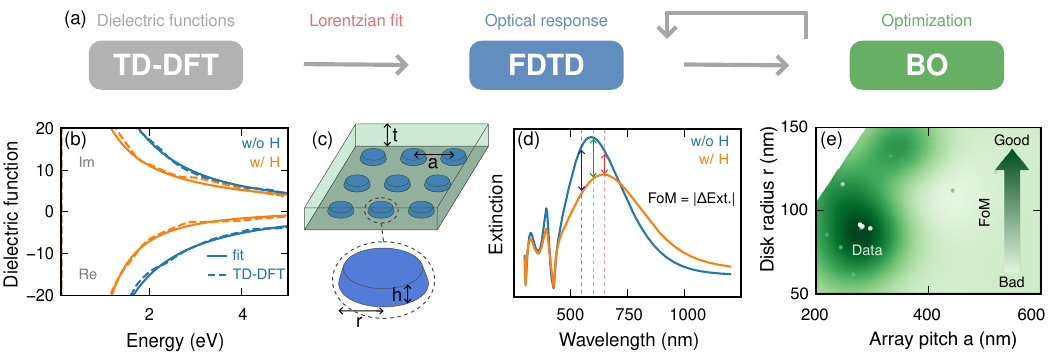}
\caption{
    \textbf{Overview of the optimization procedure.}
    (a) Flowchart of the computational methods involved.
    (b) Calculated and fitted \acrlong{df} before and after hydrogenation used to simulate the extinction spectra.
    (c) Schematic illustration of the sample geometry including the involved design parameters corresponding to disk radius $r$ and height $h$, array pitch $a$, \gls{pmma} layer thickness $t$, and Au concentration in the PdAu alloy.
    (d) Simulated extinction spectra for a nanodisk array before and after hydrogenation, used to calculate the sensing \gls{fom} from the extinction shift at a selected wavelength.
    (e) The predicted \gls{fom} landscape from the \acrshort{bo} surrogate model, used to iteratively suggest new geometries.
}
\label{fig:overview}
\glsreset{fom}
\end{figure*}

Pd is, however, associated with several challenges in the context of efficient \ce{H2} sensing \cite{DarNugLan21}, including hysteresis during H uptake and release \cite{FlaOat91, WadNugLid15}, CO poisoning \cite{DarNugLid19}, slow kinetics \cite{MarKleThe25}, low-quality localized surface plasmon resonances \cite{MarNunBod20, EkbRahRos22, NugBaiDar22}, and poor sensitivity in humid environments \cite{TheAndNil25}.
These challenges have motivated extensive research over the past decade, resulting in mitigation strategies such as alloying \cite{NugDarZhd18, DarNugLid19, TheAndNil25}, applying coatings \cite{HonLeeSeo15, NugDarCus19, DarStoOst20}, nanostructuring \cite{NugBaiDar22}, and machine learning-based signal processing \cite{MarKleThe25}.
With these efforts, Pd-based nanomaterials have evolved into a highly tunable platform for plasmonic H sensing.
Current research efforts are, however, typically limited to trial-and-error approaches targeting each challenge in isolation, and a unifying perspective that systematically connects the thermodynamics, kinetics, optical response, and practical device constraints remains absent.

To this end, we propose here a flexible computational framework to model and optimize sensor systems.
We focus on sensors based on PdAu nanodisks due to the hysteresis reduction and improved kinetics during H sorption associated with Au  \cite{LuoWanFla10,WadNugLid15}.
The nanodisks are further organized into ordered 2D arrays, or metasurfaces, leading to narrower optical resonances, known as \glspl{slr}, arising from their collective optical response \cite{HumBar14, NugBaiDar22}.
This phenomenon requires a uniform refractive index surrounding the nanodisks, achieved by coating the arrays with \gls{pmma}, which matches the refractive index of the silica substrate while having a limited impact on \ce{H2} absorption rates in the nanodisks \cite{NugDarCus19}.
The optical response of the alloy material is described by its macroscopic \gls{df}, obtained from first-principles density functional theory calculations, as a function of the Au and H concentrations.
The sensor system is characterized based on continuum-scale electrodynamic simulations of the extinction spectrum before and after hydrogenation.

This framework is deployed in \gls{bo} to optimize sensor design with respect to the five-dimensional parameter space spanned by nanodisk height and radius, array pitch (disk center-to-center distance), \gls{pmma} layer thickness, and Au concentration in the alloy.
\Gls{bo} is an iterative process where a surrogate model is used to predict an objective function in the entire domain while quantifying the associated uncertainty.
This guides the selection of points to sample by balancing exploration of unsampled regions and exploitation of regions close to optima.

In the present work, the objective function is the sensor \gls{fom}, which can be defined as any function of the extinction spectrum at different levels of hydrogenation.
We explore different \glspl{fom} but mainly focus on a cost-efficient single-wavelength sensing principle by defining the \gls{fom} as the change in extinction at a selected wavelength upon hydrogenation \cite{CheTraWen17, HerSteStr20, CelYeWan21}.
Compared to typical academic approaches, where the full extinction spectrum is recorded, this represents a reduced set of observables, lowering the cost and complexity of the required setup and helping to bridge the gap between academia and industry, since single-wavelength readout is typically preferred for real sensor devices.

Furthermore, our approach takes into account the complicated thermodynamics of H absorption by explicitly including the non-linear dependence of the absorbed H concentration on the surrounding \ce{H2} partial pressure.
This is achieved by optimizing sensors for a specific target partial pressure and avoiding the Au concentration-dependent hysteresis region, enabling high sensitivity across the relevant partial pressure interval.
In the present work, we focus on the \qtyrange{1}{100}{\milli\bar} range, which covers the detection limits set by the U.S. Department of Energy corresponding to \qty{1}{\milli\bar} and \qty{40}{\milli\bar} for automotive and stationary applications, respectively \cite{DoE15}, with the latter coinciding with the lower flammability limit of \ce{H2} in air.
Kinetics and environment-dependent factors, such as CO or the role of humidity, are not explicitly considered in the present work, but the framework could be extended to multiobjective optimization and even multiplexed sensors in the future.

In the following, we first describe the computational framework, which combines first-principles \gls{df} calculations with electrodynamic simulations and \gls{bo} for sensor design.
We then present the optimization results for single-wavelength sensors targeting \ce{H2} partial pressures in the \qtyrange{1}{100}{\milli\bar} range, and show that this approach efficiently identifies experimentally feasible designs with good sensing performance.
This is followed by experimental validation of the predicted structures, where we find good overall agreement with simulations despite remaining discrepancies.
We attribute remaining deviations primarily to the accuracy of the calculated \gls{df}, and discuss how these could be systematically improved in future work.
We conclude by discussing the implications of our findings for practical sensor design and future extensions of the optimization framework.

\section*{Results and Discussion}

\begin{figure}
\centering
\includegraphics{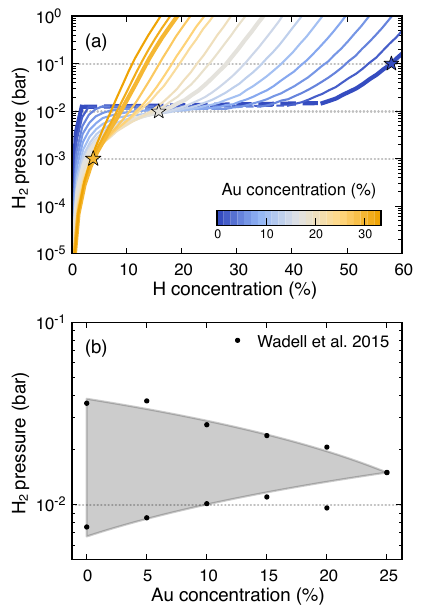}
\caption{
    \textbf{Thermodynamic model.}
    (a) H uptake pressure-composition isotherms for different alloy compositions at \qty{300}{\kelvin} from Ref.~\citenum{RahLofFra21}.
    The stars indicate the compositions of the optimized \qty{650}{\nano\meter} single-wavelength sensors at each target pressure.
    (b) Hysteresis gap as a function of alloy composition based on Ref.~\citenum{WadNugLid15} (dots) and the corresponding linear fit of the gap region limits (limits of the shaded region).
}
\label{fig:thermodynamics}
\end{figure}

The computational framework (\autoref{fig:overview}a) starts at the density functional theory-level with the set of \glspl{df} for the bulk \ce{PdAuH} system calculated in Ref.~\citenum{EkbRahRos22}, which are fitted to a Lorentzian representation (see \autoref{snote:lorentzian-rep}) to allow for smooth and continuous changes with Au and H concentrations (\autoref{fig:overview}b, \autoref{snote:lorentzian-rep}).
Each candidate sensor consists of a square periodic array of nanodisks on a silica substrate coated with \gls{pmma} (\autoref{fig:overview}c).
The nanodisks have tapered sides, with an angle of \qty{60}{\degree} between the sidewall and the base plane, reflecting the truncated cone geometry that results from physical vapor deposition through a lithographic mask.
The sensor geometry is defined by five parameters: nanodisk radius $r$ (\qtyrange{50}{150}{\nano\meter}), height $h$ (\qtyrange{20}{80}{\nano\meter}), array pitch $a$ (\qtyrange{200}{600}{\nano\meter}), \gls{pmma} thickness $t$ (\qtyrange{0}{300}{\nano\meter}), and Au concentration $c_\text{Au}$ (\qtyrange{0}{30}{\percent}) in the PdAu alloy.

The extinction spectra of each configuration before and after hydrogenation to the target \ce{H2} partial pressure (this mapping is explained below) are obtained from \gls{fdtd} simulations, and the sensor performance is quantified through the selected \gls{fom} (\autoref{fig:overview}d).
Here, we focus on a single-wavelength \gls{fom} defined as the absolute change in extinction at a selected wavelength, and specifically 450, 550, or \qty{650}{\nano\meter}, chosen to represent wavelengths of commercially available blue, green, and red lasers.
We also consider an alternative full-spectrum \gls{fom} defined as the shift in the plasmon peak upon hydrogenation divided by the full width at half maximum (measured before hydrogenation).

\Gls{bo} \cite{BroCorFre10} is employed to efficiently explore the resulting five-dimensional design space with the goal of maximizing the \gls{fom}.
Here, a Gaussian process surrogate model predicts the \gls{fom} landscape based on the available data, starting from an initial set of randomly selected data points (\autoref{fig:overview}e).
In each iteration, new candidate configurations are selected by maximizing an acquisition function that balances exploitation of current knowledge about the optimum with exploration of regions of high model uncertainty (\autoref{sfig:bayesian-optimization}).
In practice, the optimization is run for 30 iterations, after which the \gls{fom} typically shows only marginal further improvement (\autoref{sfig:bayesian-optimization}c).
This is followed by five iterations of pure exploitation to ensure that the predicted maximum is verified by simulations.
The best-performing configuration is then selected for experimental validation.

The H concentration of the hydrogenated state, $c_\text{H}$, is determined from thermodynamic data from Ref.~\citenum{RahLofFra21}, which relates the target \ce{H2} partial pressure to the equilibrium absorbed bulk H concentration of the nanodisks (\autoref{fig:thermodynamics}a).
The hysteresis region as a function of Au concentration (shaded region in \autoref{fig:thermodynamics}b) is estimated based on data from Ref.~\citenum{WadNugLid15} and excluded during the optimization, effectively setting a target pressure-dependent lower limit for the Au concentration.
For instance, at \qty{10}{\milli\bar} \ce{H2}, the Au concentration needs to be at least \qty{10}{\percent}, as lower concentrations are associated with hysteresis at this pressure.
In addition to spanning established detection targets \cite{DoE15}, the chosen target pressures of 1, 10, and \qty{100}{\milli\bar} represent distinct thermodynamic regimes on both sides of the plateau pressure with and without hysteresis, enabling benchmarking across the practically relevant pressure range.

\begin{figure}
\centering
\includegraphics{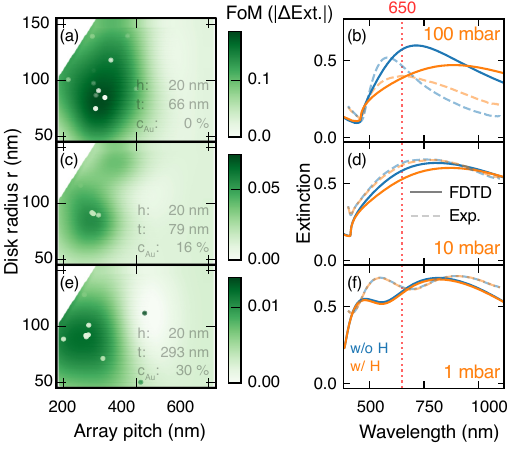}
\caption{
    \textbf{Optimization of the @\qty{650}{\nano\meter} single-wavelength sensors.}
    Two-dimensional slices of the final predicted \gls{fom} landscape for optimization at target pressures 100 (a), 10 (c) and \qty{1}{\milli \bar} (e), with the values of the other design parameters annotated.
    The corresponding extinction spectra (b, d, f) from simulations, as well as from experimental measurements, on the fabricated samples.
}
\label{fig:E-optimization}
\end{figure}

\subsection{Single-wavelength sensor @\qty{650}{\nano\meter}}

To illustrate the optimization procedure, we first present results for sensors optimized for extinction shift at \qty{650}{\nano\meter} and target pressures of 1, 10, and \qty{100}{\milli\bar}, noting that similar results are obtained at 450 and \qty{550}{\nano\meter}, as discussed in the following section.
The final predicted \gls{fom} landscapes show a single broad maximum for all target pressures (\autoref{fig:E-optimization}a, c, e).
The locations of the maxima are similar across target pressures, except for the Au concentration (see annotations in \autoref{fig:E-optimization}a, c, e), which increases as the target pressure decreases; this is expected since the H absorption increases with Au concentration below the plateau pressure (\autoref{fig:thermodynamics}a).
The \gls{pmma} thickness varies between the optimized samples, but this is due to an overall weak effect of the \gls{pmma} thickness on performance rather than a strong preference (\autoref{sfig:bayesian-optimization}).

The design parameters at the predicted optima are rounded to facilitate fabrication (\autoref{stab:samples}) and the resulting sensor geometries are fabricated and their response to pulses of \ce{H2} is measured.
The overall spectral shapes show qualitative agreement between experiment and simulation, though systematic shifts of the peaks are observed in the measurements (\autoref{fig:E-optimization}b, d, f).
These deviations can be attributed to two main factors.
First, the fabrication process is associated with a noticeable uncertainty in the final size parameters (\autoref{sfig:size-correction}), in particular due to the thermal annealing necessary to induce alloy formation \cite{NugIanWag16}, which commonly reduces the aspect ratio and causes blueshifts of the plasmon peak \cite{EkbRahRos22}.
Second, the first-principles-based \glspl{df} of the plasmonic material are subject to the inaccuracies associated with the underlying exchange-correlation functional in the density functional theory calculations \cite{RahTibRos20}, as well as the phenomenological treatment of the Drude peak \cite{gpaw}, which is fundamentally sample-dependent and would differ between an idealized material and a sample with defects.
In addition, the magnitude of the H-induced optical shift is smaller in the experiments than in the simulations (see, for instance, \autoref{fig:E-optimization}d), which alludes to an overestimation of the H absorption in the thermodynamic model.

\begin{figure}
\centering
\includegraphics{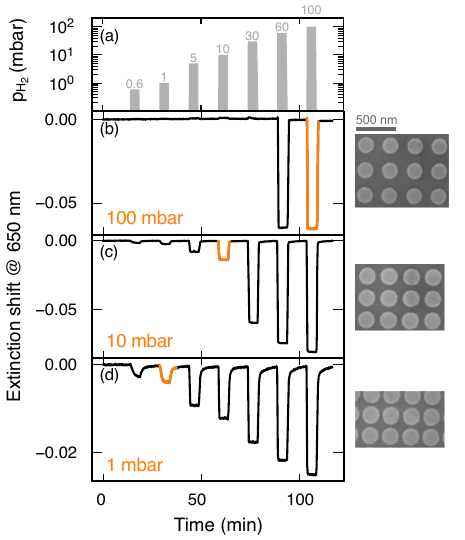}
\caption{
    \textbf{Time-resolved optical response of the @\qty{650}{\nano \meter} single-wavelength sensors.}
    (a) Sequence of hydrogen pulses ranging from \qtyrange{0.6}{100}{\milli\bar} used to test the optical extinction change at a wavelength of \qty{650}{\nano\meter} for the sensors optimized for the (b) 100, (c) 10, and (d) \qty{1}{\milli\bar} target pressures.
    The insets to the right show scanning electron microscope images of the samples.
}
\label{fig:E-performance}
\end{figure}

To fully validate the performance of the fabricated sensors, each sample was exposed to a sequence of \ce{H2} pulses with partial pressures ranging from \qtyrange{0.6}{100}{\milli\bar} (\autoref{fig:E-performance}a).
We find that, as intended, the sensors have a clear and reversible response at their target pressure.
The \qty{10}{\milli\bar} sensor performs well over the entire pressure range, whereas the \qty{100}{\milli\bar} sensor has almost no signal in the lower pressure interval.
The \qty{1}{\milli\bar} sensor, while having an overall clear signal across the pressure range, shows worse kinetics and a lower amplitude optical readout compared to the other sensors.
These results confirm that the optimized sensors realize high sensitivity across a broad range of pressures down to \qty{0.6}{\milli\bar} \ce{H2} partial pressure.

\begin{figure}
\centering
\includegraphics{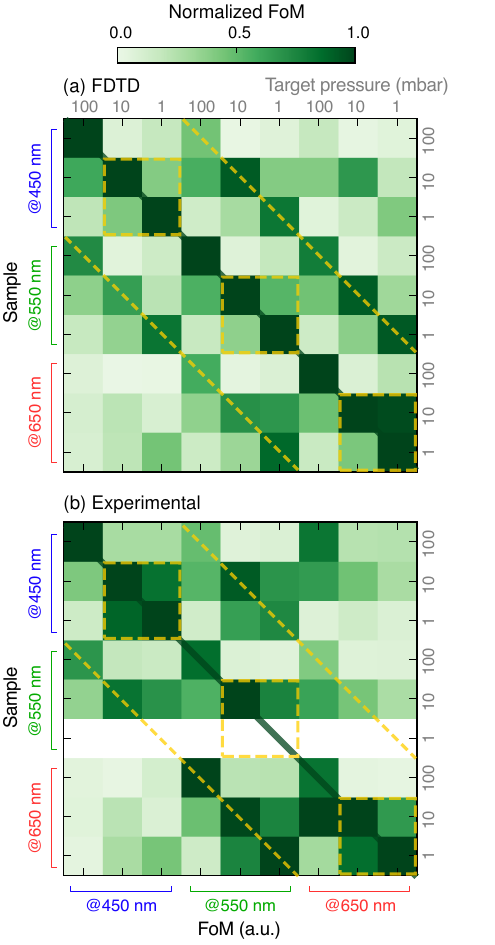}
\caption{
    \textbf{\Gls{fom} comparison for all single-wavelength sensors.}
    All considered \glspl{fom} for all samples from the simulations (a) and experimental verification (b).
    The lower x-axis indicates the \gls{fom} and the upper x-axis further specifies at which target pressure the \gls{fom} is evaluated.
    The y-axis corresponds to the samples ordered such that their respective target \gls{fom} follows the same order as the x-axis.
    The darkest color indicates the highest \gls{fom} in each column, which is expected to coincide with the dark green diagonal, while the yellow dashed lines highlight secondary observed features.
    Note that the sample corresponding to the sixth row of the lower panel is missing due to fabrication issues.
}
\label{fig:FoM-map}
\end{figure}

\subsection{All single-wavelength sensors compared}

Sensors were also optimized for single-wavelength shifts at 450 and \qty{550}{\nano\meter} with similar results (\autoref{sfig:single-wl}).
The predicted \gls{fom} landscapes are similar across all wavelengths, as well as target pressures, with a slight shift of the global maxima toward smaller array pitch and disk radius with decreasing wavelength, consistent with the size-dependent blueshift of the plasmonic resonance.
The weak dependence on the \gls{pmma} thickness remains for all single-wavelength sensors, with the optimized \gls{pmma} thickness spanning the full investigated range without a systematic trend.
The primary function of the \gls{pmma} layer is to enable the formation of \glspl{slr}.
However, the extinction spectra of the best-performing samples exhibit broad peaks near the operational wavelength, characteristic of localized surface plasmon resonance behavior rather than the narrow features typically associated with \glspl{slr}.
This observation explains the limited impact of the \gls{pmma} thickness on the optimization results.
All optimal configurations feature densely packed arrays, which can be attributed to the higher effective plasmonic surface area enhancing the absolute extinction and, consequently, the magnitude of the \ce{H2}-induced optical shift.
While the initial motivation for using ordered rather than random arrays was that the resulting \glspl{slr} would provide narrower optical features and thus improved sensing performance, these results suggest that, for these \glspl{fom}, the main advantage actually arises from the higher achievable packing density.

In \autoref{fig:FoM-map}a we plot the normalized theoretical \glspl{fom} of all single-wavelength sensors investigated in this study.
We find a significant increase in performance across the diagonal, indicating the success of our approach at tailoring sensor performance to specific pressure ranges.
In addition, a pattern of additional diagonal lines (yellow dashed lines) shifted by three steps appears.
This feature corresponds to a large extinction shift at a wavelength $\pm \qty{100}{\nano\meter}$ from the targeted wavelength, at the same target pressure, which is expected due to the broad plasmon peaks of these systems.
Similarly, there is a correlation between the sensors optimized for the two lower target pressures (yellow dashed squares), where sensors optimized for \qty{10}{\milli \bar} typically perform well for \qty{1}{\milli \bar} and vice versa.
This correlation likely emerges because, at the Au concentrations considered here, both pressures fall in similar thermodynamic regimes and thus exhibit similar \ce{H2} absorption behavior (\autoref{fig:thermodynamics}a).

The same comparison for the experimentally fabricated and measured samples shows mostly similar behavior (\autoref{fig:FoM-map}b) with six out of the eight fabricated samples being the best performing sensor at their intended target.
Due to the aforementioned discrepancy between the simulations and experimental measurements, some deviations from the simulation-based results are evident.
Most notably, the off-diagonal to the left (yellow dashed line in \autoref{fig:FoM-map}b) is very prominent for several samples, indicating good performance at a wavelength \qty{100}{\nano\meter} lower than the target.
This is consistent with two effects observed in many of the fabricated samples relative to the simulations: a blueshift of the main plasmon peak and an increase in relative amplitude of peaks at shorter wavelengths with respect to those at longer wavelengths (\autoref{sfig:single-wl}b).
Both of these effects yield a larger extinction at the \qty{100}{\nano\meter} lower wavelength, leading to an increased \gls{fom}.
In addition, the correlation between the sensors at the lower target pressures (yellow dashed squares) predicted theoretically is also visible in the measured response.

Lastly, we have also optimized sensors geometries for the same target pressures but for a \gls{fom} based on the full spectral readout, corresponding to the H-induced peak shift divided by the full width at half maximum (\autoref{sfig:full-spectrum}).
This approach shows a large theoretical improvement in performance, but the global maxima are much more narrow compared to the single-wavelength sensors, making the experimental verification more difficult.
As a result, the full-spectrum sensors display a poor signal-to-noise ratio (\autoref{sfig:time-series-B}) and are unexpectedly outperformed by one of the single-wavelength sensors (\autoref{sfig:FoM-comparison-all}).
This highlights that there is considerable potential for future development if the agreement between experiment and modeling can be improved further.

\section*{Conclusions}

We have established a \gls{bo} framework for the inverse design of PdAu metasurface \ce{H2} sensors that integrates optical simulations with first-principles-based \glspl{df} for PdAu alloys in the pristine and hydrogenated states.
Applied to sensors targeting defined \ce{H2} pressures, it successfully identifies high-performance regions in the design space.
These devices meet the U.S. Department of Energy \cite{DoE15} requirements for the limit of detection using only a single-wavelength readout, demonstrating that computationally guided design can translate directly to practical, manufacturable sensors.
This approach could be extended to design multiplexed sensors, where different regions of the sensor surface target different pressure regions, and multiobjective optimization where, for instance, response kinetics, gas selectivity, or stability in humid conditions, are treated simultaneously, provided that the corresponding physical phenomena can be modeled correctly.

While the agreement between simulations and experiments is good given the first-principles basis of the \glspl{df} without any empirical adjustment, this approach is limited by the remaining discrepancies which becomes critical near narrow optima.
To reach the full potential of this approach, the sources of disagreement need to be reconciled, including (i) improved control over the geometry in the fabrication process, (ii) refinement of the \glspl{df} of the nanodisks to represent real materials, (iii) adjustment of the thermodynamic model for accurate H absorption isotherms, and (iv) refinement of the optical simulations to better describe the entire optical system.
In particular improved structural control and refined material models will be essential to fully exploit the predictive capability of this optimization framework, and will therefore have to be targeted in future work.
Once these limitations are overcome, optimization frameworks of this kind will eliminate tedious trial-and-error and grid searches associated with traditional sensor development and enable efficient and rapid design across increasingly complex parameter spaces and objective functions.

\section*{Methods}

\subsection*{Bayesian optimization}
We use \gls{bo} \cite{BroCorFre10} to iteratively optimize the \gls{fom} as a function of nanodisk height and radius, array pitch and \gls{pmma} thickness.
We consider different \glspl{fom}, which are always calculated from extinction spectra with and without H in relation to the target pressure \textit{via} \gls{fdtd} simulations.
The optimization relies on an underlying Gaussian process which predicts the mean and variance of the \gls{fom} landscape in each step.
We use the open source python package  \textsc{GPy} \cite{gpy} to set up Gaussian processes.
The covariance function consists of radial basis function kernels, one for each optimization parameter (\textit{i.e.,} a total of four), with $\Gamma$ functions as hyperpriors for the length scale and variance of the kernels.

The \gls{fdtd} simulations are the by far the most time consuming step.
In order to maximize efficiency, two sensor geometries are suggested in each iteration and run in parallel.
The geometries are suggested based on maximizing an acquisition function that balances exploration and exploitation.
We use the separable natural evolution strategy \cite{snes} as a global optimization algorithm to maximize an upper confidence bound acquisition function \cite{BroCorFre10}, with a constraint that ensures that the second geometry picked in every iteration is not within same hyper-rectangle surrounding the first (there is no such restriction between iterations).

\subsection*{Lorentzian fit of \glspl{df}}
The \glspl{df} used in this work are based on the set of \ce{Pd_{1-x}Au_xH_y} \glspl{df} calculated in Ref.~\citenum{EkbRahRos22} with time-dependent density functional theory.
The original calculations are performed with 24 (metal) atom unit cells which limits the composition resolution to $1/24\approx\qty{4}{\percent}$.
In order to allow for continuous changes of the \gls{df} with composition, the \gls{df} is fitted to a Lorentzian representation parametrized by polynomials of the compositions which results in smoother \glspl{df}, seemingly in better agreement with previous experimental work (\autoref{sfig:DF-fit-single-disk-sensitivity}).
Details of this procedure can be found in \autoref{snote:lorentzian-rep}.

\subsection*{FDTD simulations}
\label{sect:fdtd}
Extinction spectra were calculated using the \gls{fdtd} method implemented in Ansys Lumerical FDTD \cite{lumerical}.
The computational cell consists of a truncated cone \ce{Pd_{1-x}Au_xH_y} nanodisk placed on a silica substrate with refractive index \num{1.46}.
The nanodisk is embedded in a \gls{pmma} layer with refractive index of \num{1.51} based on experimental measurements \cite{NugAlbAnt20}.
To simulate an ordered array, periodic boundary conditions are applied in the $x$ and $y$-directions while a so-called perfectly matched layer is used in the $z$-direction.
The sample is illuminated by a plane wave source in the wavelength interval \qtyrange{250}{1250}{\nano\meter} at normal incidennce.
The hydrogen induced lattice expansion is neglected to avoid optical shifts caused by how the nanodisk is represented on the grid.

\subsection*{Sample fabrication}

Nanodisk arrays were fabricated on \qty{100}{\milli\meter} fused silica wafers (Corning 7980, \qty{500+-20}{\micro\meter} thickness, double-side polished, Ra < \qty{1}{\nano\meter}).
A bilayer electron beam lithography resist stack was prepared by spin-coating AR-P 6200 (1:2 in anisole, \qty{4000}{\rpm}, \qty{5}{\minute} bake at \qty{180}{\degreeCelsius}) followed by \gls{pmma} 950k A2 (\qty{4000}{\rpm}, identical bake).
A \qty{20}{\nano\meter} Cr discharge/reflective layer was thermally evaporated (Lesker NANO36) prior to exposure.

Patterning was performed by electron beam lithography (Raith EBPG 5200; \qty{10}{\nano\meter} beam step size, \qty{20}{\nano\ampere}).
Proximity-effect correction was applied using a Monte Carlo–based dose modulation routine.
After exposure, the Cr layer was removed by a \qty{40}{\second} Cr wet etch.

The resist stack was developed in two steps: \qty{1}{\minute} in \ce{H2O}:isopropyl alcohol (1:4) for \gls{pmma} development, followed by \qty{1}{\minute} in o-xylene for AR-P 6200.
A brief \ce{O2} plasma descum (\qty{5}{\second}, \qty{50}{\watt}, \qty{0.333}{\milli\bar} (=\qty{250}{\milli\torr}), \qty{80}{\sccm} \ce{O2}; Plasma-Therm Batchtop) was carried out prior to metallization.
Metal deposition was performed in a Lesker PVD 225, followed by overnight lift-off in Remover 1165 and sequential rinsing (acetone, isopropyl alcohol, de-ionized water).

For dicing, samples were protected with S1813 (\qty{2000}{\rpm}, \qty{1}{\minute} bake at \qty{110}{\degreeCelsius}) and diced using a DISCO DAD3350 with a resin-bonded diamond blade (K010-600JXS; \qty{2}{\milli\meter\per\second} feed, \qty{25000}{\rpm}).
The protective coating was removed by two acetone baths (\qty{5}{\minute} each) and subsequent isopropyl alcohol/de-ionized water rinsing.

\subsection*{Experimental measurements}

The setup used for evaluating the performance of the samples consists of a quartz tube plug-flow reactor with optical access for transmittance measurements (Insplorion AB), and is equipped with a set of mass flow controllers (Bronkhorst High-Tech B.V.) that control the flow rate and gas composition (argon, \ce{H2}) at atmospheric pressure.
The reactor temperature was controlled using a closed-loop temperature control system (Eurotherm 3216), with the sample surface temperature inside the chamber (measured \textit{via} a K-type thermocouple) continuously used as the input.
The chamber can accommodate up to two samples, which are illuminated using an unpolarized halogen white light source (AvaLight-HAL, Avantes) coupled through a bifurcated optical fiber (FCB-UV600-2, Avantes BV) equipped with collimating lenses.
The transmitted light from each sample is collected and analyzed by a dual channel fiber-coupled fixed-grating spectrometer (AvaSpec-ULS2048CL-2-EVO, Avantes BV).

The samples were exposed to a constant gas flow rate of \qty{200}{\milli\liter\per\minute} of argon/\ce{H2} mixtures throughout the entire measurement (at atmospheric pressure).
The process started with an initialization stage that lasted almost \qty{3}{\hour}, where the sensor temperature was set to \qty{120}{\degreeCelsius}.
During this time, the sensor was exposed ten times to \qty{100}{\milli\bar} \ce{H2} in argon, in order to (i) desorb previously adsorbed \ce{H2O} from the sample surface and (ii) acquire a stable sensor baseline.
The main stage of the experiments was performed at \qty{30}{\degreeCelsius} and consisted of three identical sets of \ce{H2} pulses.
The partial pressures of \ce{H2} in each set were \num{0.6}, \num{1}, \num{5}, \num{10}, \num{30}, \num{60}, and \qty{100}{\milli\bar}.
Within the set, each \ce{H2} pulse lasted for \qty{300}{\second}, followed by \qty{600}{\second} at \qty{1}{\bar} of argon gas pressure.

\section*{Acknowledgments}
We gratefully acknowledge funding from the Swedish Research Council (Nos.~2020-04935 and 2021-05072), the Swedish Foundation for Strategic Research (SIP21-0032), the Area of Advance Nano at Chalmers, and the Competence Centre TechForH2.
The Competence Centre TechForH2 is hosted by Chalmers University of Technology and is financially supported by the Swedish Energy Agency (P2021-90268) and the member companies Volvo, Scania, Siemens Energy, GKN Aerospace, PowerCell, Oxeon, RISE, Stena Rederi AB, Johnson Matthey, and Insplorion.
We are also grateful for computational resources provided by the National Academic Infrastructure for Supercomputing in Sweden at NSC, PDC, and C3SE partially funded by the Swedish Research Council through grant agreement No.~2022-06725, the Berzelius resource provided by the Knut and Alice Wallenberg Foundation at NSC, and the Dutch National supercomputer Snellius hosted by SURF.
Part of this work was carried out at the Chalmers MC2 cleanroom facility.

\section*{Supporting Information}
Supporting Information Available: Details on the Lorentzian fitting of the \glspl{df}; additional figures on Bayesian optimization convergence, size correction, single-wavelength and full-spectrum sensor optimization, and time-series optical response data.

\end{document}